\documentclass[aps,prc,twocolumn,showpacs,superscriptaddress,footinbib]{revtex4-1}

\usepackage{graphicx}
\usepackage{dcolumn}
\usepackage{bm}
\usepackage{enumerate}
\usepackage{amsmath}

\usepackage[normalem]{ulem}  

\begin{document}


\title{Many-particle correlations and Coulomb effects on temperatures from fragment momentum fluctuations}

\author{S.R. Souza}
\affiliation{Instituto de F\'\i sica, Universidade Federal do Rio de Janeiro Cidade Universit\'aria, \\CP 68528, 21941-972, Rio de Janeiro, Brazil}
\affiliation{Instituto de F\'\i sica, Universidade Federal da Bahia,\\
Campus Universit\'ario de Ondina, 40210-340, Salvador, Brazil}
\author{M.B. Tsang}
\affiliation{National Superconducting Cyclotron Laboratory and Department of Physics and Astronomy Department,\\ Michigan State University, East Lansing, Michigan 48824, USA}
\author{R. Donangelo}
\affiliation{Instituto de F\'\i sica, Universidade Federal do Rio de Janeiro Cidade Universit\'aria, \\CP 68528, 21941-972, Rio de Janeiro, Brazil}
\affiliation{Instituto de F\'\i sica, Facultad de Ingenier\'\i a, Universidad de la Rep\'ublica, Julio Herrera y Reissig 565, 11.300 Montevideo, Uruguay}

\date{\today}

\begin{abstract}
We investigate correlations in the
fragment momentum distribution due to the propagation of fragments under the influence of their mutual Coulomb field, after the breakup of an excited nuclear source.
The magnitude of the effects on the nuclear temperatures obtained from such distributions is estimated with the help of a simple approach in which a charged fragment interacts with a homogeneous charged sphere.
The results are used to correct the temperatures obtained from the asymptotic momentum distributions of fragments produced by a Monte-Carlo simulation in which
the system's configuration at breakup is provided by the canonical version of the Statistical Multifragmentation Model.
In a separate calculation, the dynamics of this many-particle charged system is followed in a molecular dynamics calculation until the fragments are far away from the breakup volume.
The results suggest that, although the magnitude of the corrections is similar in both models, many-particle correlations present in the second approach are non-negligible and should be taken into 
account in order to minimize ambiguities in such studies.
\end{abstract}

\maketitle

\begin{section}{Introduction}
\label{sect:introduction}
Under the assumption that a nuclear system which disassembles into many fragments
has reached thermal equilibrium at the freeze-out configuration, the determination of its temperature at this moment plays a central role in understanding the system's properties.
Indeed, it is needed to construct the basic thermodynamic quantities, such as the Helmholtz free energy, from which one may obtain relevant information as, for instance, the nuclear caloric curve and the nuclear equation of state.
It has, therefore, been studied both theoretically \cite {temperaturesBondordf1989,temperaturesBSB1989,temperatureShlomo,temperaturesKonrad1989,PochodzallaReview1997,thermometry2000,
temperaturesTrautmann2007,temperatureReviewMSU1994,BorderiePhaseTransition2008,ISMMlong,internalTemperatures2015} and
experimentally \cite{reviewSubal2001,thermometryVient2006,BorderiePhaseTransition2008,temperaturesKonrad1989,temperatureReviewMSU1994,
reviewTempeatureNatowitz,PochodzallaReview1997,temperaturesHuang1997,thermometry2000,temperaturesTrautmann2007,ccGSI,ccNatowitzHarm,
ccMa1997,ISMMlong,momentumTemperature2010} by many groups over the last decades.
Nevertheless, several difficulties have been found in these studies (see \cite{BorderiePhaseTransition2008,temperaturesTrautmann2007,temperatureReviewMSU1994,internalTemperatures2015} and references therein), leading to ambiguities and conflicting conclusions.

Recently, a method to determine the nuclear temperature $T$ at the freeze-out configuration, based on the momentum distribution of the fragments, has been proposed and used to investigate the caloric curve obtained from fragments produced in different reactions \cite{momentumTemperature2010}.
More specifically, it has been demonstrated \cite{momentumTemperature2010} that the variance $\sigma^2$ of the distribution of $q=p_x^2-p_y^2$, where $p_x$ and $p_y$ are the $x$ and $y$ Cartesian components of the fragment's momentum $\vec{p}$, is related to $T$ through:

\begin{equation}
\sigma^2=\int d^3\vec{p}\,(p_x^2-p_y^2)^2f(\vec{p})=4A^2m_n^2T^2\;,
\label{eq:sigmaT}
\end{equation}

\noindent
since $\langle q\rangle$ vanishes if $f(\vec{p})$ symbolizes the fragment's momentum distribution in the source's reference frame and it is assumed that $f(\vec{p})$ is given by the Boltzmann distribution. (A quantum treatment has also been considered in Ref.\ \cite{,momentumTemperature2012}.)
In this expression, $A$ denotes the fragment's mass number and $m_n$ is the nucleon mass.
Thus, $\sigma^2$ may be obtained experimentally for some selected species from the measured fragments' momenta.
This allows one to determine $T$ from different thermometers.

However, when applied to actual data analysis \cite{momentumTemperature2010}, the measured temperatures turned out to be much larger than those obtained in previous studies, using different methods.
Since, as mentioned above, the latter have their own limitations and uncertainties, one may not, a priori, discard the present method due to this discrepancy.
On the other hand, the elucidation of the underlying mechanisms which lead to it is important in order to improve our understanding of the multifragment emission process.

Since the charged fragments produced at breakup propagate under the influence of their mutual Coulomb field until they reach the detectors, one should investigate the extent to which the momentum distribution is affected by this process and the magnitude of its effect to the measured temperatures.
This point has already been addressed in Refs.\ \cite{CoulombCorrectionsThermometers2013,CoulombCorrectionsThermometers2014} where corrections based on a picture similar to that assumed in our deterministic model discussed below have been applied to their model calculations \cite{CoulombCorrectionsThermometers2013}.
The corresponding corrections turned out to be non-negligible.

In the present work we address this point using two different approaches.
In one of them, the canonical version of the Statistical Multifragmentation Model (SMM) \cite{smm1,smm2,smm4} is used to generate fragments, at a given temperature $T$ \cite{grandCanonicalBotvina1987,smmIsobaric}, on an event by event basis.
These fragments are then placed inside the breakup volume and allowed to propagate until they are sufficiently separated from each other, so that the Coulomb energy of the system is negligible.
The second approach is based on a simplified picture, in which a given fragment is placed inside a homogeneous charged sphere and it is then accelerated away from the breakup volume, until it is infinitely far away from the sphere.
This provides a simple deterministic manner to calculate the fragment's momentum distribution and, from it, the nuclear temperature through Eq.\ (\ref{eq:sigmaT}).
The effects estimated from this simple model are then used to correct the temperatures obtained from our Monte-Carlo simulation.
It turns out that many-particle correlations are non-negligible and, therefore, detailed treatments need to be employed in order to obtain meaningful results from such analyses.
 
The remainder of the manuscript is organized as follows: The models just mentioned are explained in Sect.\ \ref{sect:model}.
They are used to calculate the nuclear temperatures from the momentum distributions in Sect.\ \ref{sect:results}.
We conclude in Sect.\ \ref{sect:conclusions} with the main findings of this work.

 \end{section}
 
\begin{section}{The models}
\label{sect:model}
In both treatments used in this work it is assumed that a nuclear source with mass and atomic numbers $A_0$ and $Z_0$ has expanded to a  breakup volume $V_{bk}=(1+\chi)V_0$.
The parameter $V_0$ is the volume of the source at normal nuclear density and $\chi>0$ is a model parameter, which is assumed to be $\chi=2$ throughout this study, as it plays a minor role on the investigated effects.
We also use $A_0=95$ and $Z_0=45$, which corresponds to 70\% of the $^{58}$Ni+$^{78}$Kr system, studied in Ref.\ \cite{momentumTemperature2010}.
The removal of 30\% of the system aims at taking into account particles that left the source region due to the pre-equilbrium emission. 

Our Monte-Carlo simulation uses the canonical implementation of SMM \cite{grandCanonicalBotvina1987,smmIsobaric} to generate the $M$ fragments, whose multiplicity and composition of species vary for each fragmentation mode $f$.
The statistical weight of the latter is given by:

\begin{equation}
w_f=\exp\left(-F_f(T,V_{bk})/T\right)\;,
\label{eq:wcanonical}
\end{equation}

\noindent
where $F_f(T,V_{bk})$ is the Helmholtz free energy associated with the fragmentation mode \cite{smmIsobaric}.
For each event, the fragments' momenta are assigned according to the Metropolis simulation method.
More specifically, for each fragment $i$, a partner $j\ne i$ is randomly selected and a random increment $\vec{\Delta}=(\Delta_x,\Delta_y,\Delta_z)$ is sampled, where $\Delta_x$, $\Delta_y$, and $\Delta_z$ are independent random numbers.
Then, the trial momenta $\vec{p}_i^{\,t}=\vec{p}_i+\vec{\Delta}$ and $\vec{p}_j^{\,t}=\vec{p}_j-\vec{\Delta}$ are calculated.
Since the system is initialized with $\{\vec{p}_i=0\}$, this procedure ensures that the center of mass of the source remains at rest.
Then, the trial momenta are accepted with probability:

\begin{equation}
{\cal P}={\rm Min}\left\{1,\exp\left(-\frac{1}{T}\left[\frac{(p_i^t)^2-p_i^2}{2m_nA_i}+\frac{(p_j^t)^2-p_j^2}{2m_nA_j}\right]\right)\right\}\;,
\label{eq:Metropolis}
\end{equation}

\noindent
where $A_i$ ($A_j$) denotes the mass number of the {\it i}-th ({\it j}-th) fragment.
In the simulation, $10^5$ of such steps per fragment are carried out in order to generate the initial momentum distribution for a given fragmentation mode.
Although the constraint $\sum_{i=1}^{M_f}\vec{p}_i=0$ slightly distorts the momentum distribution, the Boltzmann expression 

\begin{equation}
f(\vec{p})=C\exp\left(-\frac{p^2}{2m_nAT}\right)\;,
\label{eq:fp0}
\end{equation}

\noindent
where $C$ is a normalization factor, remains an excellent approximation to it, as is illustrated in the top panel of Fig. \ref{fig:momDist}, which displays the density of particles $N(p)$ with momentum between $p$ and $p+\Delta p$, obtained with the above procedure, for protons and $^4$He's at $T=5$ MeV.

\begin{figure}[tb]
\includegraphics[width=8.5cm,angle=0]{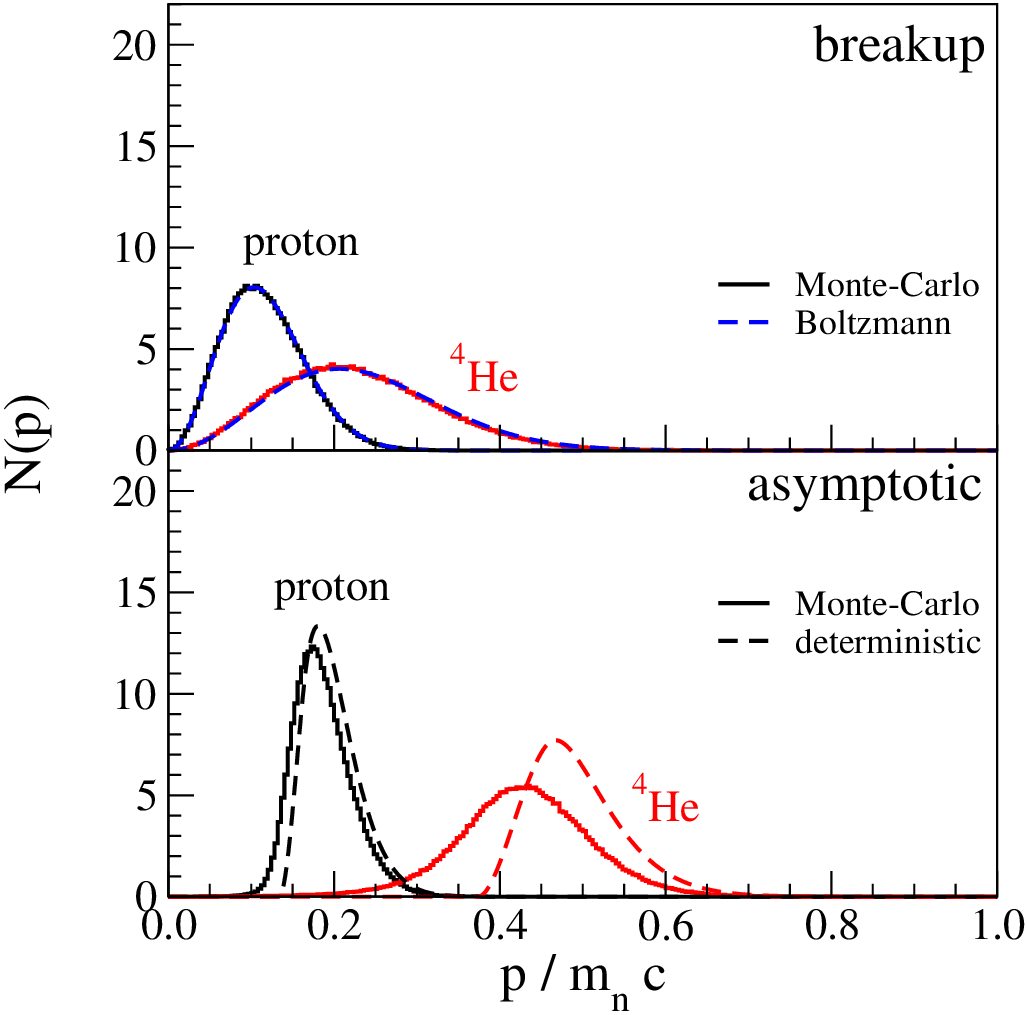}
\caption{\label{fig:momDist} (Color online) Density of particles with momentum between $p$ and $p+\Delta p$, for $T=5$ MeV.
In the top panel the Monte-Carlo distribution at breakup is displayed whereas the bottom panel exhibits the asymptotic distributions for both treatments presented in this work. For details see the text.}
\end{figure}

Next, the fragments are randomly placed inside the breakup volume using a procedure similar to that employed above.
More precisely, we initially set the position of each fragment $\vec{r}_i$ to zero, i.e. $\{\vec{r}_i=0\}$, and calculate a trial move for two fragments $i$ and $j$ according to $\vec{r}_i^{\,t}=\vec{r}_i+\vec{\Delta}\frac{A_j}{A_i+A_j}$ and $\vec{r}_j^{\,t}=\vec{r}_j-\vec{\Delta}\frac{A_i}{A_i+A_j}$, where  $\vec{\Delta}=(\Delta_x,\Delta_y,\Delta_z)$ are independent random variables.
The move is rejected if at least one of the fragments steps outside the breakup volume.
As previously, this procedure is repeated $10^5$ times for each fragment $i$.

Once the initial positions and momenta of the fragments have been initialized, their trajectories are followed in time using a standard classical molecular dynamics treatment, in which the fragments interact with each other only through the Coulomb force.
In order to speed up the numerical calculations, we adopt the Runge-Kutta-Cash-Karp method \cite{RungeKuttaCashKarp}, which automatically adjusts the time step in order to preserve a pre-established accuracy.
The fragments are then allowed to propagate away from the breakup volume until the separation among them is large enough to allow the Coulomb energy to be neglected.

This entire procedure is repeated for each event produced by SMM and the density of particles corresponding to the $i$-th species with asymptotic momentum $p_i$ between  $p$ and $p+\Delta p$ is given by:

\begin{equation}
N(p)=\frac{1}{\Delta p}\left[\sum\limits_{\substack{f \\ p \le p_i < p+\Delta p}} w_f(i)\right]/\left[\sum_f w_f(i)\right]\;,
\label{eq:npamc}
\end{equation}

\noindent
where $w_f(i)$ represents the statistical weight associated with the partition $f$ containing the fragment.
This Monte-Carlo method is used to generate many millions of partitions.

The asymptotic momentum distributions for protons and $^4$He are displayed in the bottom panel of Fig.\ \ref{fig:momDist}.  
It reveals that the distribution is shifted to higher values, leaving a hole in the region of small momentum values.
This effect appreciably affects the distribution of $q$ and, consequently, the temperatures obtained from it.
We shall come back to this point in the next section.

We now consider  a uniformly charged sphere of volume $V_{bk}=4\pi R_0^3/3$ and charge $(Z_0-Z_i)e$, where $e$ stands for the elementary charge.
This sphere remains frozen in this configuration and is kept fixed at the origin of the coordinate frame.
If a particle of charge $Z_ie$ and mass $m_i=m_n A_i$ is placed at a distance $r < R_0$ from its center, energy conservation allows one to easily calculate the particle's asymptotic momentum: $p_i^2=p_{i,0}^2+\delta^2(1-x^2/3)$, where $p_{i,0}$ is the thermal momentum at breakup, $\delta^2=3 m_i Z_i(Z_0-Z_i)e^2/R_0$, and $x\equiv r/R$.
Obviously, some of the assumptions made in this model are very unrealistic.
Nevertheless, it is intended to provide a rough estimate of the Coulomb distortions considered in this work.
However, the results presented below reveal that, in spite of this, the model gives quite reasonable results.

Upon assuming that the particle may be found with equal probability anywhere within the breakup volume, the density of particles with asymptotic momentum between $p$ and $p+\Delta p$ is given by:

\begin{equation}
N(p)=C'\int_{x_0}^1dx\,x^2 u(p,x)\exp\left(-\frac{u(p,x)}{2m_nAT}\right)\;,
\label{eq:npasa}
\end{equation}

\noindent
where $u(p,x)\equiv p^2-\delta^2(1-x^2/3)$, $C'$ is a normalization factor and $x_0={\rm Min}(1,\sqrt{{\rm Max}(0,3[1-p^2/\delta^2])})$.
The above integral may be carried out analytically and the result reads:

\begin{equation}
N(p)=C'[G(p,1)-G(p,x_0)]\;,
\label{eq:npasa2}
\end{equation}

\noindent
where

\begin{eqnarray}
&&G(p,x)=\Big\{\exp\left(-\frac{\delta^2x^2}{6m_nAT}\right)2x\delta\sqrt{m_nAT}\Big[-3(p^2-\delta^2)\nonumber\\
&& -x^2\delta^2-9m_nAT\Big]+{\rm erf}\left(\frac{x\delta}{\sqrt{6m_nAT}}\right)3\sqrt{6\pi m_n A T}\nonumber\\
&& \Big[p^2-\delta^2+3m_nAT\Big]\Big\}\exp\left(-\frac{p^2-\delta^2}{2m_nAT}\right)\frac{m_n A T}{2\delta^3}\;.
\label{eq:G}
\end{eqnarray}

\noindent
The distribution $N(p)$ predicted by this model is also displayed in the Bottom panel of Fig.\ \ref{fig:momDist}.
Comparison between these results with those obtained with the Monte-Carlo treatment just presented shows that, although there are noticeable differences between them, this simple model reproduces the main qualitative features, such as the suppression of particles in the small momentum region.
Furthermore, since the width and position of the maximum of the distributions given by the two models do not differ by a large factor, one should expect them to predict temperatures of the same magnitude.
Therefore, this simple model may provide useful estimates, although discrepancies should be expected as the distributions differ.

It is easy to understand the reasons for the observed differences.
In the Monte-Carlo treatment, the fragments interact with each other as they travel away from the breakup region.
Therefore, the assumption that a fragment interacts with a static homogeneous sphere, made in our deterministic model, is a crude representation of the actual scenario.
For instance, a fragment may eventually be left behind by all the others if its initial velocity is much smaller than theirs.
In this case, it would feel a very weak Coulomb force, leading to the appearance of a smooth tail in the region of small momentum.
On the other hand, in the deterministic model, no particle may have energy smaller than that corresponding to the Coulomb energy at the surface of the breakup volume, which leads to the steep fall off observed in the bottom panel of Fig.\ \ref{fig:momDist}.

\end{section}

\begin{section}{Results}
\label{sect:results}
We now turn to the distribution of $q$ and to the temperature determination.
We construct the distribution $N_q$, in the following manner.
First, the $q$ axis is divided into bins of width $\Delta q$.
Then, for each particle of the selected species $i$, found in one event of the Monte-Carlo model, the statistical weight $w_f$ of the partition is added to $N_q$, so that:

\begin{equation}
N_q=\frac{1}{\Delta q}\left[\sum\limits_{\substack{f \\ q \le q_i < q+\Delta q}} w_f(i)\right]/\left[\sum_f w_f(i)\right]\;,
\label{eq:nqxymc}
\end{equation}

\noindent
where $q_i\equiv p_{i,x}^2-p_{i,y}^2$.

The model predictions are exhibited in the top panel of Fig.\ \ref{fig:qDist}, for the $^4$He nucleus, at breakup (full lines) and at the asymptotic configuration (dashed lines).
The breakup temperature is $T=5$ MeV.
These results clearly reveal that the Coulomb interaction in the late stages of the process leads to an important broadening of the initial $q$ distribution, which should result in larger reconstructed temperatures, as we show below.

\begin{figure}[tb]
\includegraphics[width=8.5cm,angle=0]{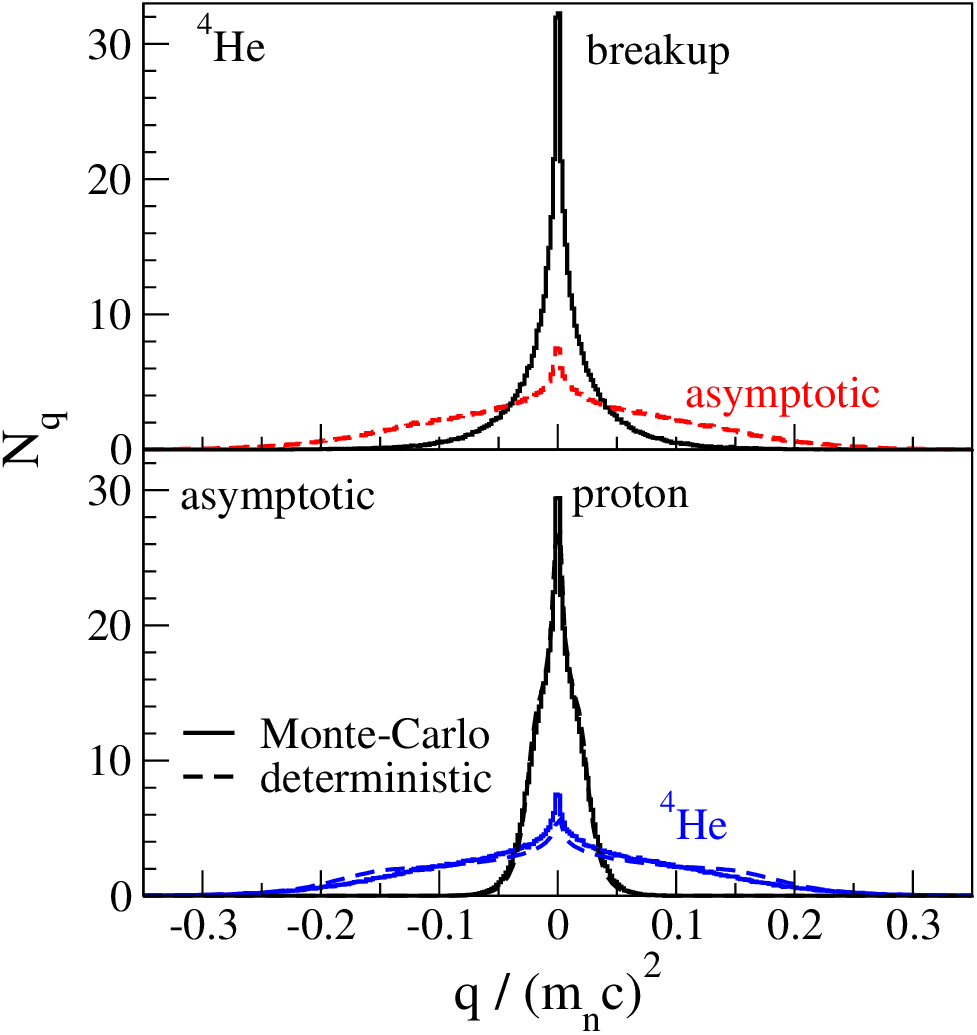}
\caption{\label{fig:qDist} (Color online) Distribution of $q$ for $T=5 MeV$.
Top panel: Results predicted by the Monte-Carlo model at the breakup stage (full line) and after the propagation in the Coulomb field (dashed line).
The selected species is the $^4$He nucleus.
Bottom panel: Asymptotic $q$ distributions for proton and $^4$He. The full lines correspond to the results obtained with the Monte-Carlo model whereas the dashed lines represent the predictions made with the deterministic model.
For details see the text.}
\end{figure}

To calculate the distribution $N_q$ with the deterministic model, the $x$ and $y$ momentum axes are discretized, so that $p_{i,j}=\sqrt{p_{x_i}^2+p_{y_j}^2+p_z^2}$, where $p_{x_i}$ and $p_{y_j}$ are the discrete components of the momentum vector.
In this way, $N_q$ reads:

\begin{equation}
N_q=\frac{1}{\Delta q}\left[\sum\limits_{\substack{i,j \\ q \le q_{i,j} < q+\Delta q}} {\cal N}_{i,j}(p_{i,j})\right]/\left[\sum_{i,j} {\cal N}_{i,j}(p_{i,j})\right]\;,
\label{eq:nqxyd}
\end{equation}

\noindent
where

\begin{equation}
{\cal N}(p_{i,j})=\int_{-\infty}^{+\infty} dp_z \left[H(p_{i,j},1)-H(p_{i,j},x_0)\right]\;,
\label{eq:nqxydd}
\end{equation}

\begin{eqnarray}
&&H(p_{i,j},x)=\int_{x_0}^1 dx\, x^2\exp\left(-\frac{u(p_{i,j},x)}{2m_nAT}\right)\nonumber\\
&=&\frac{3m_nAT}{2\delta^3}\exp\left(-\frac{p_{i,j}^2-\delta^2}{2m_nAT}\right)\Big\{-2x\delta\exp\left(-\frac{\delta^2x^2}{6m_nAT}\right)\nonumber\\
&&+\sqrt{6\pi m_nAT}\,{\rm erf}\left(\frac{x\delta}{\sqrt{6m_nAT}}\right)\Big\}\;,
\label{eq:Hnqxyd}
\end{eqnarray}

\noindent
and  $q_{i,j}=p_{x_i}^2-p_{y_j}^2$.

The bottom panel of Fig.\ \ref{fig:qDist} shows the comparison between the asymptotic $q$ distribution calculated with the two models, for proton and $^4$He, produced at the break temperature $T=5$ MeV.
As anticipated, the models make similar predictions, despite being based on different scenarios for the propagation in the Coulomb field.
However, non-negligible discrepancies are observed, which must lead to different reconstructed temperatures. 

The variance of the distribution is calculated through:

\begin{equation}
\sigma^2=\left[\sum\limits_{\substack{f \\ q \le q_i < q+\Delta q}} q_i^2 w_f(i)\right]/\left[\sum_f w_f(i)\right]
\label{eq:variancemc}
\end{equation}

\noindent
in the case of the Monte-Carlo model.
In the deterministic treatment, it is given by

\begin{equation}
\sigma^2 =\left[\sum\limits_{\substack{i,j \\ q \le q_{i,j} < q+\Delta q}} q_{i,j}^2{\cal N}_{i,j}(p_{i,j})\right]/\left[\sum_{i,j} {\cal N}_{i,j}(p_{i,j})\right]\;.
\label{eq:varianced}
\end{equation}

\noindent
From the above expressions and Eq. (\ref{eq:sigmaT}), the temperature may be calculated for each species.

\begin{figure}[tb]
\includegraphics[width=8.5cm,angle=0]{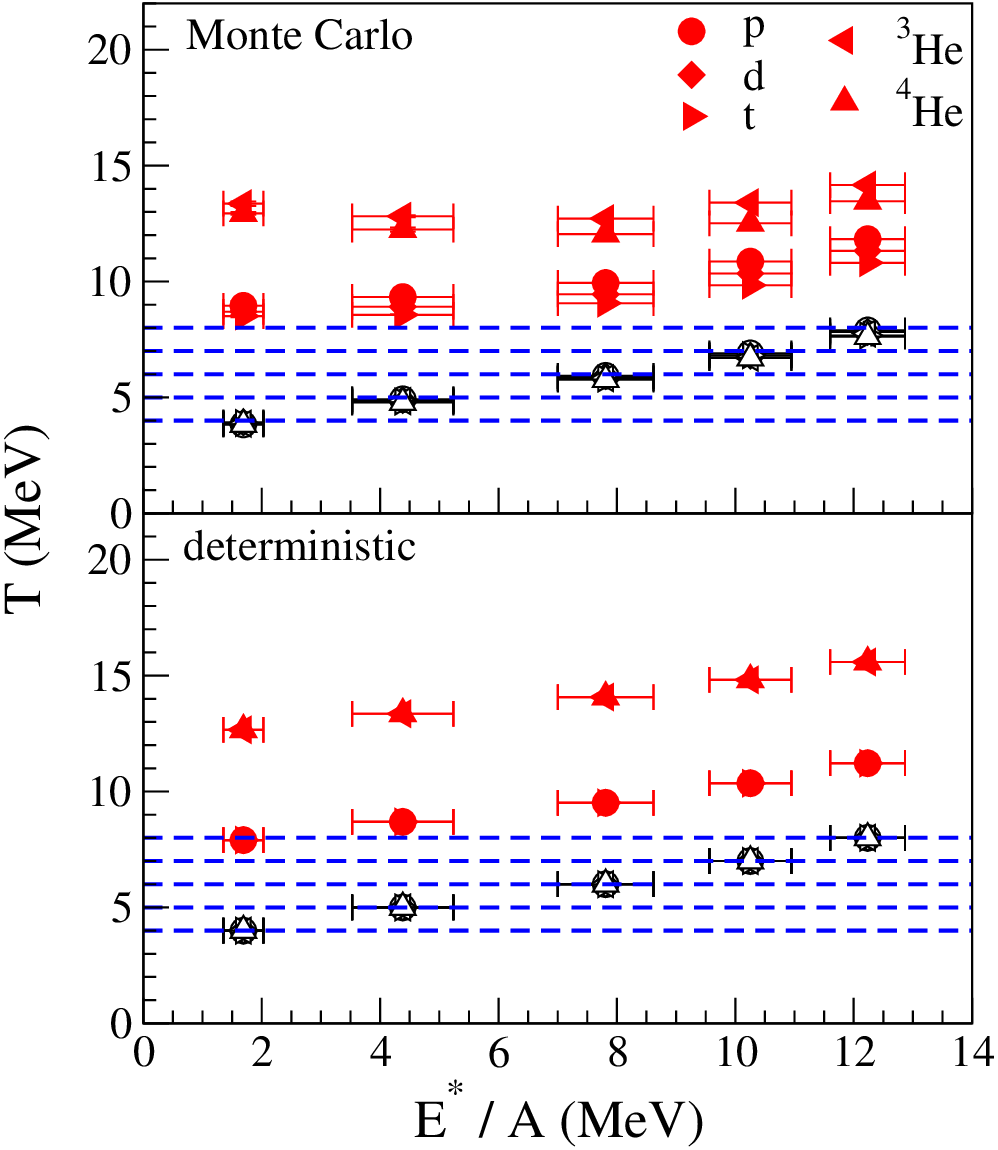}
\caption{\label{fig:ccComp} (Color online) Caloric curves constructed from temperatures associated with different species.
The predictions of the Monte Carlo model are displayed in the top panel whereas those from the deterministic model are exhibited in the bottom panel.
The dashed lines represent the input temperatures used in the SMM calculations, which also provide both the average excitation energy and its width.
The open symbols represent the calculations at breakup and the full ones correspond to the asymptotic values.
For details see the text.}
\end{figure}

In order to construct the caloric curve, the excitation energy of the source at a given breakup temperature $T$ is provided by SMM, as well as the corresponding width.
The caloric curves predicted by the two models are displayed in Fig.\ \ref{fig:ccComp} for temperatures obtained with different species.
The top panel shows the results calculated with the Monte-Carlo Model whereas those predicted by the deterministic model are shown in the bottom panel.
The open symbols correspond to the temperatures at breakup.
By construction, the predictions of both models agree with the input temperatures, represented by the horizontal dashed lines.
This scenario is completely changed when the asymptotic distributions are used.
Indeed, in agreement with the results obtained in Ref.\ \cite{CoulombCorrectionsThermometers2013}, the reconstructed temperatures are substantially larger than the input ones.
In both treatments, the fragments with larger charges are more strongly affected due to the larger Coulomb forces that act upon them.

As expected from the above results, the temperatures predicted by our two models are of the same magnitude, in spite of some qualitative differences.
First, it may be noted a weak $A$ dependence of the temperatures reconstructed from the Monte-Carlo treatment, whereas this effect is negligible in our deterministic model.
This $A$ dependence is due to distortions of the momentum distribution due to the center of mass constraints.
The most noticeable difference is observed in the behavior of $T$ as a function of the excitation energy $E^*$.
The deterministic model exhibits a monotonous increase as a function of $E^*$, whereas the Monte-Carlo model predicts that $T$ first decreases at low excitation energies and then increases for $E\gtrsim 6$ MeV.
By discarding half of the events and recalculating the temperatures, we have checked that this unexpected behavior is not due to the statistics.
The changes turned out to be smaller than the symbols' sizes.
The explanation must be related to the low fragment multiplicity values at low input temperatures.
This leads to strong correlations between the fragments' momenta due to the constraint $\sum_i\vec{p}_i=0$.
The average multiplicities are $2\pm 1$, $5.5\pm 1.8$, $10.5\pm 2.1$, $13.5\pm 2.1$, and $15.4\pm 2.1$ for $T= 4$, 5, 6, 7, and 8 MeV, respectively.
As the multiplicity increases, the effect on the individual fragments' momenta becomes weaker.

\begin{figure}[tb]
\includegraphics[width=8.5cm,angle=0]{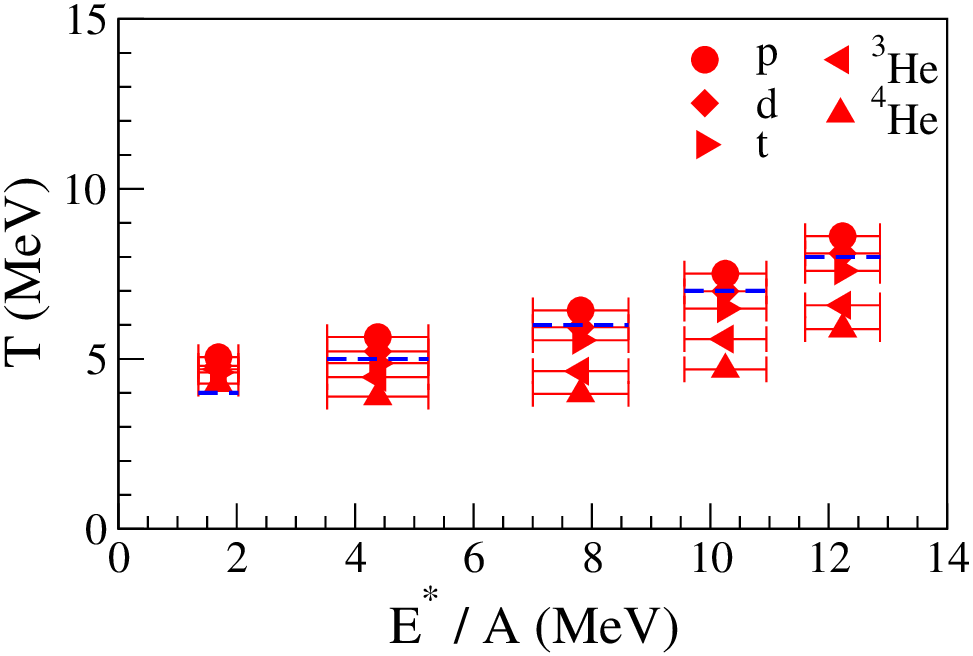}
\caption{\label{fig:ccCorect} (Color online) Caloric curves constructed using temperatures given by the Monte-Carlo treatment, corrected by the deterministic model.
The dashed lines represent the input temperatures used in the SMM calculations.
For details see the text.}
\end{figure}

The systematic enhancement of the temperatures due to the Coulomb boost is tentatively removed from the Monte-Carlo results by subtracting from them the difference between the asymptotic and the breakup temperatures given by the deterministic model.
The results are exhibited in Fig.\ \ref{fig:ccCorect}.
Although the disagreement with the input-values (dashed lines) is substantially reduced, the results clearly show that the method should not be applied at low excitation energies.
They also reveal that the discrepancies are larger for fragments with larger atomic number.
Furthermore, the observed dispersion of the values obtained with different species suggests that their use may lead to ambiguities in determining the nuclear caloric curve.
Our results seem to indicate that detailed Monte-Carlo treatments should be used in data analysis in order to minimize the difficulties just mentioned.

Obviously, the magnitude of the reported changes will be weakened if the source's charge is reduced.
We have checked, using the deterministic model, that the effects are slightly reduced if we assume that 50\%, instead of 30\%, of the mass and charge are emitted in the pre-equilibrium emission.
However, the changes are not large enough to modify our conclusions.
Finally, we have not varied the breakup volume as its influence should be very small, as long as reasonable values are used, since the initial Coulomb energy is proportional to the radius, which varies very slowly with the volume.

As a final remark, the deexcitation of the primary excited fragments has been entirely disregarded in our calculations, although its contribution to the final yields is important at the temperatures considered in this work \cite{ISMMlong,nuclearThermometry2000}.
Nevertheless, the consideration of this aspect will not affect our conclusions and, therefore, for the sake of simplicity, it has been not included in our treatments, although it should be taken into account in actual data analysis.

\end{section}

\begin{section}{Concluding Remarks}
\label{sect:conclusions}
We have investigated the influence of many-particle correlations and of the Coulomb interaction in the late stages of the dynamics on the fragments' momentum distribution, which are used to reconstruct the temperatures at breakup.
Using a Monte-Carlo treatment, in which the partition mode and the system's temperature are provided by SMM, and another model in which a fragment interacts with a static homogeneous charged sphere, we calculated the fragments' momenta when they are very far away from the breakup volume.
In agreement with previous results \cite{CoulombCorrectionsThermometers2013}, we have found that the Coulomb repulsion leads to important distortions on the fragments' momentum distribution.
It leads to an appreciable enhancement of the reconstructed temperatures.
By correcting the temperatures predicted by our Monte-Carlo model, subtracting from them the enhancement observed in the deterministic model, the agreement with the input temperatures has been significantly improved.
However, many particle effects turn out to be important and may lead to conflicting conclusions if the method is applied to low excitation energy values or if different species are used as thermometers.
We therefore suggest that a detailed Monte-Carlo treatment to eliminate these effects should be developed in order to reliably obtain breakup temperatures from such measurements.

\end{section}

\begin{acknowledgments}
We would like to thank Dr. A. McIntosh for fruitful discussions.
This work was supported in part by CNPq, FAPERJ BBP grant, and the National Science Foundation under Grant No. PHY-1102511.
We also thank the
Programa de Desarrollo de las Ciencias B\'asicas (PEDECIBA) and the
Agencia Nacional de Investigaci\'on e Innovaci\'on (ANII) for partial financial support.
\end{acknowledgments}

\bibliography{manuscript}
\bibliographystyle{apsrev4-1}

\end{document}